\documentclass[prl,twocolumn,english,showpacs,floatfix,superscriptaddress]{revtex4-1}
\usepackage{graphicx}
\usepackage{amssymb}
\usepackage{color}
\usepackage{amsmath}

\begin{document}
\title{Probing the bond order wave phase transitions of the ionic Hubbard model by superlattice modulation spectroscopy}

\author{Karla Loida}
\affiliation{HISKP, University of Bonn, Nussallee 14-16, 53115 Bonn, Germany}
\author{Jean-S\'ebastien Bernier}
\affiliation{HISKP, University of Bonn, Nussallee 14-16, 53115 Bonn, Germany}
\author{Roberta Citro}
\affiliation{Dipartimento di Fisica “E.R. Caianiello“, Universit\`a degli Studi di Salerno, Via Giovanni Paolo II 132, I-84084 Fisciano (Sa), Italy}
\author{Edmond Orignac}
\affiliation{Universit\'e de Lyon, \'Ecole Normale Sup\'erieure de Lyon, Universit\'e Claude Bernard, CNRS, Laboratoire de Physique, F-69342 Lyon, France}
\author{Corinna Kollath}
\affiliation{HISKP, University of Bonn, Nussallee 14-16, 53115 Bonn, Germany}

\begin{abstract}
An exotic phase, the bond order wave, characterized by the spontaneous dimerization of the hopping,
has been predicted to exist sandwiched between the band and Mott insulators in
systems described by the ionic Hubbard model.
Despite growing theoretical evidences, this phase still evades
experimental detection. Given the recent realization of the ionic Hubbard model
in ultracold atomic gases, we propose here to detect the bond order wave
using superlattice modulation spectroscopy. We demonstrate, with the help of time-dependent density-matrix
renormalization group and bosonization, that this spectroscopic approach
reveals characteristics of both the Ising and Kosterlitz-Thouless transitions signaling the presence of
the bond order wave phase. This scheme also provides insights into the excitation spectra
of both the band and Mott insulators.
\end{abstract}

\date{\today}
\maketitle
%
In solid state materials, the combination of strong interactions, quantum
fluctuations and finely tuned energy scales gives rise to rich physics.
For example, in a large class of materials including transition metal
oxides~\cite{ImadaTokura1998}, organics~\cite{PowellMcKenzie2011} and
iridates~\cite{WitczakKrempaBalents2014}, the presence of strong
on-site repulsion between fermions leads to
the suppression of charge motion and to the formation of Mott insulating states.
Inducing charge fluctuations around these
Mott insulators, e.g.~by doping, reveals intricate phase diagrams highlighting
the presence of multiple competing orders.
Perhaps one of the best known examples is the
emergence of $d$-wave superconductivity in high-temperature cuprates at the interface
between antiferromagnetic and Fermi liquid phases~\cite{BednorzMuelller1986}.

Complex states also arise near phase transitions when competing insulating
effects are present. In the neighborhood of such transitions, where the strength of the
insulating terms is comparable, the effect of smaller terms, such as the kinetic energy,
leads to the emergence of metallic phases or exotic correlations. For example, at the
interface between the Mott and band insulators, such a competition is believed to play an
important role in the ionic to neutral transitions in organic charge-transfer
solids~\cite{TorranceLaPlaca1981,NagaosaTakimoto1986} and at ferroelectric
transitions in perovskites~\cite{EgamiTachiki1993}. The ionic Hubbard model, which gained
prominence over the last decade, was first developed to explain the physics near these
transitions. In this model, the on-site Hubbard repulsion and staggered potential terms induce
insulating behavior when taken separately,
but when taken together they can compete and give rise
to a region of increased charge fluctuations.
This region, occuring where these two
terms are of comparable strength, is of great interest as this competition
leads to the emergence of the bond order wave phase signaled by a spontaneous dimerization of
the hopping.

\begin{figure*}
\includegraphics[width=0.8\linewidth,clip=true]{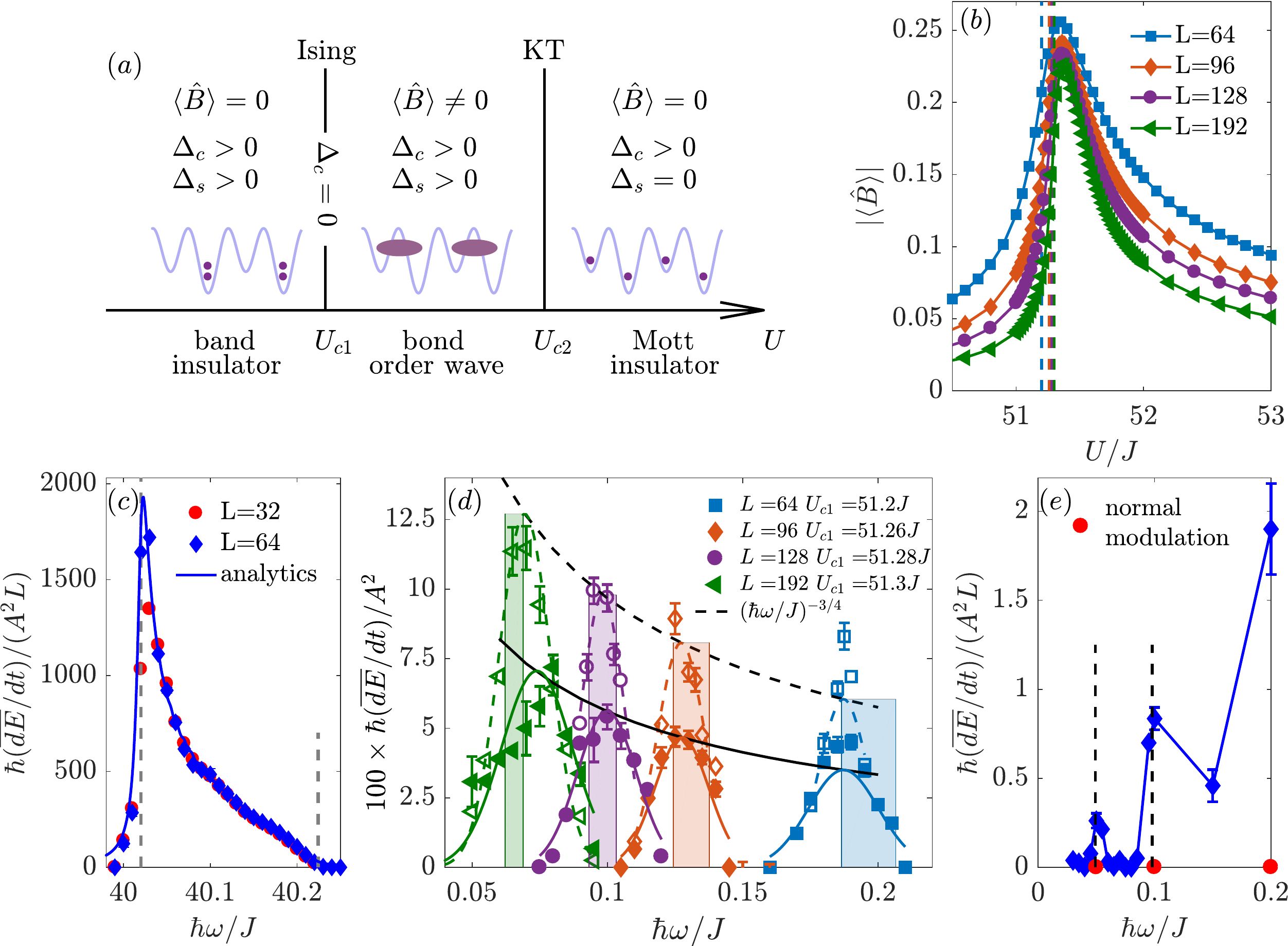}
\caption{(a) Phase diagram of the 1D ionic Hubbard model as a function of $U$ for
  fixed $\Delta$ and $J$. $\Delta_c$ and $\Delta_s$ are the charge and spin gaps.
  (b) The bond order parameter $|\langle \hat B\rangle|$ at fixed $\Delta=50J$
  for different system sizes calculated from DMRG keeping up to $400$ states.
  The dashed vertical lines mark the position of the maximum of the derivative located at $U_{c1} \sim 51.2J,\ 51.26J,\ 51.28J,\ 51.3J$
  for system sizes $L=64,\ 96,\ 128,\ 192$, respectively. (c)-(e) Energy absorption rate for different regions of the phase diagram
  at fixed $\Delta=50J$:  (c) Deep in the band insulator for $U=10J$, $A=0.001J$, dashed vertical lines mark the
  width of the band of excitations; (d) At the Ising critical point for different system sizes and amplitudes
  $A=0.001J$ (filled symbols) and $A=0.0005J$ (open symbols). The Gaussian fits (solid and dashed lines) are guides to the eye
  where the maxima are fixed to $\lambda (\hbar\omega/J)^{-3/4}$ with $\lambda$ chosen to agree with the $L=128$ peaks;
  (e) In the bond order wave phase near the KT transition for $L=64$, $U=52J$ and $A=0.005J$~\cite{footnote1}.
  Vertical dashed lines indicate multiples of the spin gap.}
\label{fig:phasediagram}
\end{figure*}
This model is described by the standard Hubbard Hamiltonian to which a
staggered potential, with an energy offset $\Delta$ between neighbouring sites,
is added
\begin{eqnarray}
  \nonumber \hat{H}_0&=&-J\sum_{j=1,\sigma}^{L-1} (c_{j\sigma}^{\dag} c_{j+1 \sigma} + \text{h.c.})
  + U\sum_{j=1}^{L}n_{j \uparrow}n_{j\downarrow} \\
  &+& \frac{\Delta}{2}\sum_{j=1,\sigma}^{L}(-1)^jn_{j \sigma}. \nonumber
\end{eqnarray}
Here $c_{j,\sigma}^{(\dag)}$ are the fermionic annihilation (creation) operators and $n_{j,\sigma}$ is
the particle number operator on site $j$ with spin $\sigma=\{\uparrow, \downarrow\}$.
The amplitude $J$ is the hopping matrix element, $U$ the repulsive on-site
interaction strength, and $L$ the number of sites. In one dimension, this model has been the subject of a large number of
studies using a variety of techniques including bozonization, density renormalization group,
exact diagonalization and quantum Monte Carlo
methods~\cite{FabrizioNersesyan1999, FabrizioNersesyan2000, GidopoulosTosatti2000, WilkensMartin2001, TorioCeccatto2001,
  KampfBrune2003, ZhangLin2003, BatistaAligia2004, ManmanaSchoenhammer2004,OtsukaNakamura2005, LegezaSolyom2006,  TincaniBaeriswyl2009}.
A smaller number of studies have also focused on the excitations of the ionic Hubbard
model~\cite{AligiaBatista2005,HafezJafari2010, GoJeon2011, HafezTorbatiUhrig2014, HafezTorbatiUhrig2015}.
Despite initial controversy, theoretical investigations point to the existence, at half filling,
of a bond order wave phase occuring around $U \sim \Delta$ characterized by the spontaneous dimerization of the
hopping, i.e.~the order parameter $B=|\langle \hat B\rangle|$,
\begin{eqnarray}
\hat B=\sum_{j=1,\sigma}^{L-1}(-1)^j\left(  c_{j \sigma}^{\dag} c_{j+1 \sigma} + \text{h.c.}\right). \nonumber
\end{eqnarray}
The bond order wave spontaneously breaks site-inversion symmetry, and, in the limit of infinite system
size, the state is two-fold degenerate with restored bond-inversion symmetry on either even or odd bonds.
The bond order wave phase possesses finite charge and spin gaps and is separated,
on the one side, from a band insulating state by an Ising quantum
phase transition and, on the other side, from a Mott insulator by a Kosterlitz-Thouless (KT)
transition. The ground state phase diagram as a function of the onsite interaction
strength, $U$, is shown in Fig. \ref{fig:phasediagram}(a). Despite strong theoretical evidence,
the bond order wave has yet to be experimentally detected.

Ultracold fermionic gases provide an appealing novel avenue to detect this state. The
ionic Hubbard model was recently realized using fermions loaded into an optical superlattice
potential~\cite{TarruellEsslinger2012}.
Furthermore, over the last decade, powerful detection techniques were developed to probe the
intricate physics of quantum gases. For instance, characteristic excitations of correlated
systems can be probed using various spectroscopic techniques such as radio frequency, Raman,
Bragg and lattice modulation spectroscopy~\cite{BlochZwerger2008,ToermaeTarruell2015}.
In particular, for fermionic systems,
modulating the lattice potential was proposed as an approach to detect the pairing gap in a
superfluid state and the spin ordering in Mott insulating states~\cite{KollathGiamarchi2006},
and as a possible thermometer~\cite{LoidaKollath2015}. Numerous theoretical studies have also
considered the response of double occupancy to the modulation of the lattice
amplitude~\cite{KollathGiamarchi2006, HuberRuegg2009,
  SensarmaDemler2009,MasselToermae2009,KorolyukToermae2010,
  XuJarrell2011,TokunoGiamarchi2012_0,TokunoGiamarchi2012,DirksFreericks2014}. Experimentally,
this response was used to investigate the fermionic Mott insulator~\cite{JoerdensEsslinger2008},
to probe nearest-neighbor correlations~\cite{GreifEsslinger2011}, and to determine the
lifetime of doublon excitations~\cite{StrohmaierDemler2010}. 
In addition, mapping out higher Bloch bands in a quasimomentum-resolved fashion~\cite{HeinzeSengstock2011}
and studying interband dynamics~\cite{HeinzeBecker2013} were also carried out using lattice modulation.
Directional lattice modulation spectroscopy was used to study charge density
wave order in the two-dimensional ionic Hubbard model on an honeycomb lattice~\cite{MesserEsslinger2015}. 

Detecting the bond order wave and characterizing
the nature of the neighboring phase transitions requires the development of a technique which couples
directly to the order parameter. We demonstrate here, using time-dependent matrix
renormalization group (t-DMRG)~\cite{DaleyVidal2004, WhiteFeiguin2004, Schollwoeck2011}
and bosonization techniques~\cite{DelfinoMussardo1998,BajnokWagner2001,TakacsWagner2006}, that {\it superlattice} amplitude
modulation spectroscopy reveals features
of both the Ising and KT transitions in one-dimensional systems signaling the presence of
the bond order wave phase. This modulation, also provides insights into the excitation spectra of both the
band and Mott insulators, the two phases bordering the bond order wave. On the band insulating
side, close to the Ising transition, and in the bond order wave, close to the KT transition,
one can follow the closing of the excitation gaps.

The proposed detection method relies on inducing a small time-periodic modulation of the superlattice
potential described by the perturbation
$\hat{H}_{\text{pert}} = A ~ \sin(\omega t) ~ \hat B$,
with $A$ the strength and $\omega$ the frequency of the modulation directly coupling to the order parameter.
By contrast with standard lattice amplitude modulation, here the lattice is modulated
in a dimerized fashion. Experimentally such a perturbation is achievable by time-periodically tuning the phase between
the two laser waves generating the optical superlattice. The number of excitations created through the modulation
is then assessed by monitoring the absorbed energy.

Using t-DMRG, we determine the evolution of the full system
$\hat{H}_0+\hat{H}_{\text{pert}}$, with open boundary conditions, and
compute the energy of the system as a function of time. Typically, the energy shows a quadratic rise at initial times
before entering a linear regime and then saturates at later times. We extract the slope of the
linear energy increase, a quantity we identify as the energy absorption rate.
For the time evolution using t-DMRG we typically keep $120$ states
(except at the Ising critical point where we keep $160$). The error analysis are performed by increasing the matrix dimension
to $160$ states (and $240$ at the Ising critical point).
The time-step of the Trotter-Suzuki time evolution is set to $J\Delta t=0.001\hbar$ and we use $J\Delta t=0.0005\hbar$
to conduct the error analysis. The error bars provided in the figures represent the maximal uncertainties due to
the matrix dimension, the time-step and the fit error (as the fit range has been varied).

In the limit of sufficiently weak modulation strength, where linear response theory applies, the energy absorption rate
is related to the imaginary part of the dynamic susceptibility associated with the order parameter of the bond
order wave, i.e.  $\overline{dE}/dt \sim \omega A^2 \text{Im}\chi_{\hat{B}\hat{B}^\dag}(\omega)$.
Here the imaginary part of the dynamic susceptibility $\chi_{\hat{B}\hat{B}^\dag}(\omega)$
is determined by the Fourier transform of the retarded correlation
function $\chi_{\hat{B}\hat{B}^\dag}(t)=-i\theta(t)\langle [\hat B(t),\hat B^\dag(0)]\rangle_0$
where $\langle \cdot \rangle_0$ denotes the expectation value with respect to the ground state.

%
We first consider the structure of the energy absorption rate under the effect of superlattice modulation
at the Ising critical point $U_{c1}$. Within bosonization, $\hat B$ is
proportional to the Ising order parameter in the vicinity of the
Ising transition~\cite{FabrizioNersesyan1999, FabrizioNersesyan2000}.
At the Ising quantum critical point, the scaling dimension of the order parameter is $1/8$. Estimating its dynamic
susceptibility at criticality by a scaling argument, we find the absorbed power to diverge as $L~\omega^{-3/4}$
in the thermodynamic limit as the modulation frequency decreases to zero signaling the Ising transition.
At frequencies lower than the spin gap, the charge fluctuations dominate causing the divergence.
One should note that normal lattice modulation fails to detect the existence of the Ising phase transition.
In this case, the energy absorption rate does not present a divergence at the Ising transition as this latter
modulation scheme does not couple to the bond order wave.

The form of the divergence is affected by the system size.  
In a finite system, the imaginary part of the dynamic susceptibility is
\begin{eqnarray}
  \text{Im}\chi_{\hat B \hat B^\dag}(\omega) \sim L^{\frac{7}{4}}~
  \sum_{n=0}^{\infty} \left(\frac{\Gamma\left( n+1/8\right)}{\Gamma\left( n+1\right)}\right)^2
  \delta \left(\hbar (\omega- \Omega(n)) \right) \nonumber
\end{eqnarray}
with $\hbar\Omega(n) = 4\pi\frac{\hbar u}{aL}(n+\frac{1}{16})$
where $a$ is the lattice spacing and $u$ the sound velocity of the low energy excitations.
Thus, for a finite system, the divergence will be signaled by the presence of a series of peaks occuring at
$\hbar\Omega(n)$ with spectral weight scaling as $\omega^{-3/4}$.

In order to test these low energy predictions, we time-evolve systems of different sizes at the Ising critical point
for a range of modulation frequencies and extract the energy absorption rates. At low energies ($\hbar\omega/J < 0.15$),
our numerical results are in good agreement with bosonization and in particular the peak height follows
well the predicted $(\hbar \omega/J)^{-3/4}$ divergence.
To identify values of $U/J$ near the Ising critical point $(U/J)_{c1}$,
we find for each system size the location of the maximum of the derivative
of the order parameter, $\partial |\langle \hat B\rangle |/\partial U$ (Fig. \ref{fig:phasediagram}(b)).
Analyzing the energy absorption rate at these values, we then identify the
$n=1$ peak~\cite{footnote2}. We estimate the peak position by fitting a Gaussian to the t-DMRG
data for the system of size $L = 128$ and modulation amplitude $A = 0.0005J$ 
finding $\hbar \omega_{\text{peak}}/J \approx 0.098 \pm 0.005$.
From this we extract the sound velocity of excitations $\frac{\hbar u}{aJ} \approx 0.94 \pm 0.05$, 
a typical value for lattice systems. Using the value of the sound velocity $u$ extracted for $L=128$,
we then determine the $n=1$ peak positions for the other system sizes which we indicate in
Fig. \ref{fig:phasediagram}(d) by shaded regions. The agreement between these peaks predicted
from bosonization and the t-DMRG ones is good. We attribute the disagreement
in the peak position and height for $L=64$ to the breakdown of bosonization in the energy range where
the corresponding $n = 1$ peak appears. While the asymmetry of the $n=1$ peak at $L=192$ and $A = 0.001J$
is due to saturation effects in the numerics.

\begin{figure}
  \includegraphics[width=.85\columnwidth,clip=true]{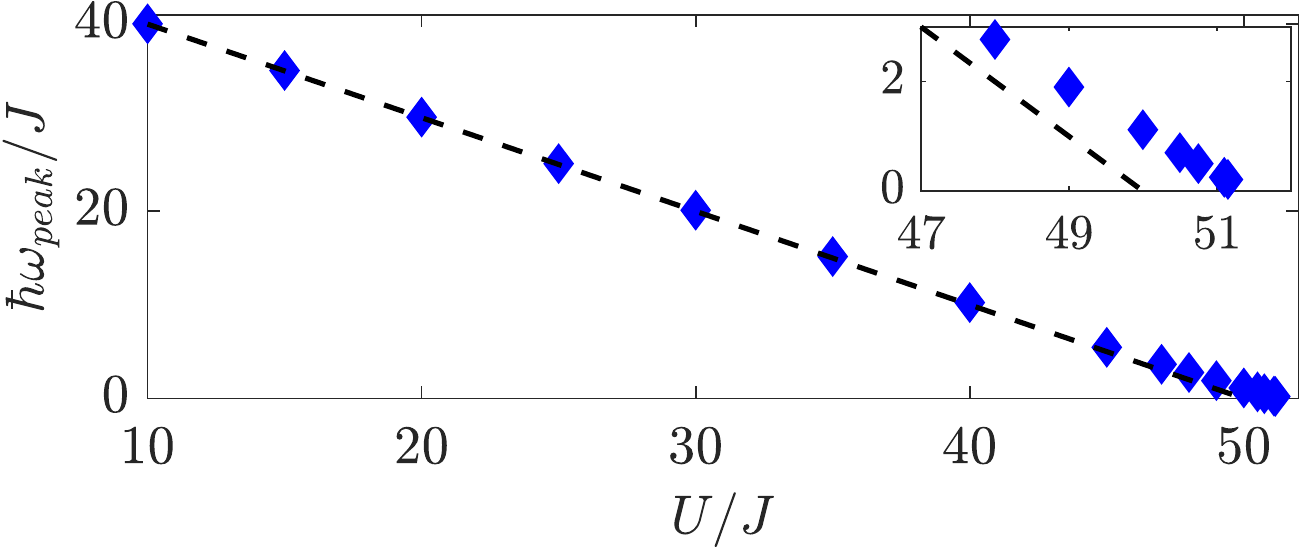}
  \caption{The position of the maximum of the absorption peak $\hbar \omega_{\text{peak}}$ corresponds approximately to the charge gap
  on the band-insulating side ($\Delta > U$) at fixed $\Delta=50J$ and $L=64$.
  Inset: close to the transition the deviation from the naive expectation $\Delta-U$ (dashed line) increases.}
\label{fig:chargegap}
\end{figure}

%
The approach to the Ising phase transition is also detected from the signal obtained on the band insulating
side as one can follow the linear closing of the charge gap
as the system approaches the critical point. In Fig. \ref{fig:phasediagram}(c),
we present the response deep within the band insulator.
The absorption peak is located approximately at $\hbar \omega \sim (\Delta-U)$, it is very sharp and its width scales
approximately as $J^2/(\Delta - U)$. A strong rise occurs at its left boundary corresponding to a
divergence at the lower excitation band edge as also seen in the non-interacting model.
The location, width and shape of the peak deep in the band insulator (see solid line in Fig. \ref{fig:phasediagram}(c))
is understood within an effective model using a Schrieffer-Wolff transformation for the excited states~\cite{inpreparation}.
When approaching the Ising transition
by increasing $U$, the peak broadens and shifts towards smaller energies. The position of the peak maximum as a
function of $U$ is shown in Fig. \ref{fig:chargegap}, thus confirming that the charge
gap closes when approaching the Ising phase transition.

%
By comparison to the Ising transition, the location of the KT quantum critical point
is harder to pinpoint in finite size systems as the bond order wave order parameter falls off very slowly, see Fig. \ref{fig:phasediagram}(b).
Nevertheless, to the left of the KT transition on the bond order wave side, bosonization predicts a
gapped response followed by a sharp increase of the energy absorption rate
at twice the minimum energy (mass) of a soliton, $\mu$, corresponding to the creation of a soliton-antisoliton pair
\begin{eqnarray}
\text{Im}\chi_{\hat B \hat B^\dag}(\omega)\sim \frac{1}{J}~\sqrt{\left ( \frac{\hbar\omega}{2\mu}\right)^2-1}~. \nonumber
\end{eqnarray}
In the spontaneous dimerized phase, singlets form either on even or odd bonds giving rise to a
doubly degenerate ground state and solitons are interpreted as domain walls between the two ground states. 
The operator $\hat B$ is SU($2$) invariant, and can only induce transitions from the singlet ground state to excited states within
the same spin sector. The lowest available state is a pair of domain walls (solitons) of opposite spins giving a threshold at
twice the soliton mass \cite{Lecheminant2005}.

To test this prediction, we time-evolve the system near this second phase boundary at $U=52J$ for a range of modulation frequencies
and extract the energy absorption rate. As shown in Fig. \ref{fig:phasediagram}(e), we observe sharp rises of the absorption rate
at multiples of the spin gap value. The spin gap is obtained from static DMRG calculations in different $S_z=N_{\uparrow}-N_{\downarrow}$
sectors, $\Delta_s=E_0(N=N_{\uparrow}+N_{\downarrow}=L,S_z=2)-E_0(N=N_{\uparrow}+N_{\downarrow}=L,S_z=0)$.
For $U=52J$ we find $\Delta_s\approx0.049J$ for $L=64$ converged in the matrix dimension. 
Note, that the normal lattice modulation does not couple to the soliton-antisoliton excitation.
Approaching the KT transition from the bond order wave side, the soliton mass becomes smaller
and smaller until at the transition the gap closes and a low energy feature, associated with
spin excitations, arises on the Mott-insulating side.
Hence, superlattice modulation succeeds in signaling the proximity of the KT quantum
critical point.
%
%
\begin{figure}
\includegraphics[width=.99\columnwidth,clip=true]{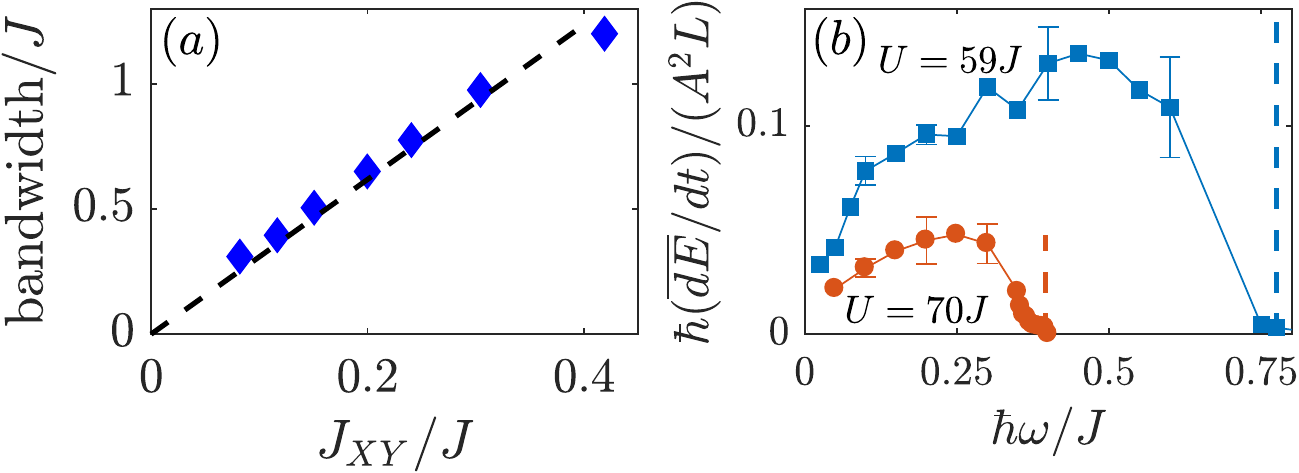}
\caption{(a) Width of the low energy band of spin excitations in the Mott insulator, extracted from the energy absorption rate, as
  a function of the effective spin coupling $J_{XY}$ for $\Delta < U$ at fixed
  $\Delta=50J$ and $L=64$, $A=0.005J$.
  The dashed line is a linear fit to the data.
  (b) Energy absorption rate as a function of $\hbar \omega$ ($\Delta=50J$, $L=64$, $A=0.005J$).}
\label{fig:bandwidth}
\end{figure}
In the Mott insulating phase, the averaged energy absorption rate is predicted by bosonization to
be constant at low modulation frequencies in the infinite size limit. In fact, in a finite system we expect equal weight
peaks which blend into a constant spectrum when $L\rightarrow\infty$.
This spectrum, whose amplitude decreases with increasing $U$, is bounded by the low energy cutoff.
Our numerical results for the case of the Mott insulator
show at low energies a broad excitation spectrum (see Fig. \ref{fig:bandwidth} (b)).  We also
find that the width and height of the spectrum decrease with increasing $U$. These different features
corroborate the predictions from bosonization. Additionally, due to the dominant role played by the
spin degrees of freedom in this region of the phase diagram, for $J \ll U-\Delta$, the ionic Hubbard
model can be mapped at low energies onto an isotropic Heisenberg chain with exchange
interaction $J_{XY}=J_Z=\frac{4J^2}{U}\frac{1}{1-(\Delta/U)^2}$~\cite{NagaosaTakimoto1986}. As seen
in Fig. \ref{fig:bandwidth}, the width of the spectrum increases linearly with the strength of the
Heisenberg exchange interaction confirming the spin nature of this excitation spectrum. 
Substructures in the spectrum might arise at longer timescales as revealed in \cite{MasselToermae2009} for
the homogeneous Hubbard model.

%
In summary, we demonstrated that superlattice modulation spectroscopy can be used
to detect features of both the Ising and KT transitions signaling the presence
of the bond order wave phase. This approach would provide a first experimental
glimpse into a phase that has evaded detection in the solid state context
and highlights the versatility of spectroscopic methods.

\paragraph{Acknowledgment}
We thank A. Sheikhan for insightful discussions. We
acknowledge financial support of the DFG (SFB 1238 project C05),
ERC Phonton (648166).


\begin{thebibliography}{50}%
\makeatletter
\providecommand \@ifxundefined [1]{%
 \@ifx{#1\undefined}
}%
\providecommand \@ifnum [1]{%
 \ifnum #1\expandafter \@firstoftwo
 \else \expandafter \@secondoftwo
 \fi
}%
\providecommand \@ifx [1]{%
 \ifx #1\expandafter \@firstoftwo
 \else \expandafter \@secondoftwo
 \fi
}%
\providecommand \natexlab [1]{#1}%
\providecommand \enquote  [1]{``#1''}%
\providecommand \bibnamefont  [1]{#1}%
\providecommand \bibfnamefont [1]{#1}%
\providecommand \citenamefont [1]{#1}%
\providecommand \href@noop [0]{\@secondoftwo}%
\providecommand \href [0]{\begingroup \@sanitize@url \@href}%
\providecommand \@href[1]{\@@startlink{#1}\@@href}%
\providecommand \@@href[1]{\endgroup#1\@@endlink}%
\providecommand \@sanitize@url [0]{\catcode `\\12\catcode `\$12\catcode
  `\&12\catcode `\#12\catcode `\^12\catcode `\_12\catcode `\%12\relax}%
\providecommand \@@startlink[1]{}%
\providecommand \@@endlink[0]{}%
\providecommand \url  [0]{\begingroup\@sanitize@url \@url }%
\providecommand \@url [1]{\endgroup\@href {#1}{\urlprefix }}%
\providecommand \urlprefix  [0]{URL }%
\providecommand \Eprint [0]{\href }%
\providecommand \doibase [0]{http://dx.doi.org/}%
\providecommand \selectlanguage [0]{\@gobble}%
\providecommand \bibinfo  [0]{\@secondoftwo}%
\providecommand \bibfield  [0]{\@secondoftwo}%
\providecommand \translation [1]{[#1]}%
\providecommand \BibitemOpen [0]{}%
\providecommand \bibitemStop [0]{}%
\providecommand \bibitemNoStop [0]{.\EOS\space}%
\providecommand \EOS [0]{\spacefactor3000\relax}%
\providecommand \BibitemShut  [1]{\csname bibitem#1\endcsname}%
\let\auto@bib@innerbib\@empty
\bibitem [{\citenamefont {Imada}\ \emph {et~al.}(1998)\citenamefont {Imada},
  \citenamefont {Fujimori},\ and\ \citenamefont {Tokura}}]{ImadaTokura1998}%
  \BibitemOpen
  \bibfield  {author} {\bibinfo {author} {\bibfnamefont {M.}~\bibnamefont
  {Imada}}, \bibinfo {author} {\bibfnamefont {A.}~\bibnamefont {Fujimori}}, \
  and\ \bibinfo {author} {\bibfnamefont {Y.}~\bibnamefont {Tokura}},\ }\href
  {\doibase 10.1103/RevModPhys.70.1039} {\bibfield  {journal} {\bibinfo
  {journal} {Rev. Mod. Phys.}\ }\textbf {\bibinfo {volume} {70}},\ \bibinfo
  {pages} {1039} (\bibinfo {year} {1998})}\BibitemShut {NoStop}%
\bibitem [{\citenamefont {Powell}\ and\ \citenamefont
  {McKenzie}(2011)}]{PowellMcKenzie2011}%
  \BibitemOpen
  \bibfield  {author} {\bibinfo {author} {\bibfnamefont {B.~J.}\ \bibnamefont
  {Powell}}\ and\ \bibinfo {author} {\bibfnamefont {R.~H.}\ \bibnamefont
  {McKenzie}},\ }\href {http://stacks.iop.org/0034-4885/74/i=5/a=056501}
  {\bibfield  {journal} {\bibinfo  {journal} {Reports on Progress in Physics}\
  }\textbf {\bibinfo {volume} {74}},\ \bibinfo {pages} {056501} (\bibinfo
  {year} {2011})}\BibitemShut {NoStop}%
\bibitem [{\citenamefont {Witczak-Krempa}\ \emph {et~al.}(2014)\citenamefont
  {Witczak-Krempa}, \citenamefont {Chen}, \citenamefont {Kim},\ and\
  \citenamefont {Balents}}]{WitczakKrempaBalents2014}%
  \BibitemOpen
  \bibfield  {author} {\bibinfo {author} {\bibfnamefont {W.}~\bibnamefont
  {Witczak-Krempa}}, \bibinfo {author} {\bibfnamefont {G.}~\bibnamefont
  {Chen}}, \bibinfo {author} {\bibfnamefont {Y.~B.}\ \bibnamefont {Kim}}, \
  and\ \bibinfo {author} {\bibfnamefont {L.}~\bibnamefont {Balents}},\ }\href
  {\doibase 10.1146/annurev-conmatphys-020911-125138} {\bibfield  {journal}
  {\bibinfo  {journal} {Annual Review of Condensed Matter Physics}\ }\textbf
  {\bibinfo {volume} {5}},\ \bibinfo {pages} {57} (\bibinfo {year} {2014})}
  \BibitemShut
      {NoStop}%
%
\bibitem{BednorzMuelller1986} J.~G. Bednorz and K.~A. Mueller, Z. Physik B - Condensed Matter {\bf 64}, 189 (1986).
%
\bibitem [{\citenamefont {Torrance}\ \emph {et~al.}(1981)\citenamefont
  {Torrance}, \citenamefont {Girlando}, \citenamefont {Mayerle}, \citenamefont
  {Crowley}, \citenamefont {Lee}, \citenamefont {Batail},\ and\ \citenamefont
  {LaPlaca}}]{TorranceLaPlaca1981}%
  \BibitemOpen
  \bibfield  {author} {\bibinfo {author} {\bibfnamefont {J.~B.}\ \bibnamefont
  {Torrance}}, \bibinfo {author} {\bibfnamefont {A.}~\bibnamefont {Girlando}},
  \bibinfo {author} {\bibfnamefont {J.~J.}\ \bibnamefont {Mayerle}}, \bibinfo
  {author} {\bibfnamefont {J.~I.}\ \bibnamefont {Crowley}}, \bibinfo {author}
  {\bibfnamefont {V.~Y.}\ \bibnamefont {Lee}}, \bibinfo {author} {\bibfnamefont
  {P.}~\bibnamefont {Batail}}, \ and\ \bibinfo {author} {\bibfnamefont {S.~J.}\
  \bibnamefont {LaPlaca}},\ }\href {\doibase 10.1103/PhysRevLett.47.1747}
  {\bibfield  {journal} {\bibinfo  {journal} {Phys. Rev. Lett.}\ }\textbf
  {\bibinfo {volume} {47}},\ \bibinfo {pages} {1747} (\bibinfo {year}
  {1981})}\BibitemShut {NoStop}%
\bibitem [{\citenamefont {Nagaosa}\ and\ \citenamefont {ichi
  Takimoto}(1986)}]{NagaosaTakimoto1986}%
  \BibitemOpen
  \bibfield  {author} {\bibinfo {author} {\bibfnamefont {N.}~\bibnamefont
      {Nagaosa}}\ and\ \bibinfo {author} {\bibfnamefont {J.}~\bibnamefont
      {Takimoto}},\ }\href {\doibase 10.1143/JPSJ.55.2735} {\bibfield  {journal}
  {\bibinfo  {journal} {Journal of the Physical Society of Japan}\ }\textbf
  {\bibinfo {volume} {55}},\ \bibinfo {pages} {2735} (\bibinfo {year}
  {1986})}
  \BibitemShut {NoStop}%
\bibitem [{\citenamefont {Egami}\ \emph {et~al.}(1993)\citenamefont {Egami},
  \citenamefont {Ishihara},\ and\ \citenamefont {Tachiki}}]{EgamiTachiki1993}%
  \BibitemOpen
  \bibfield  {author} {\bibinfo {author} {\bibfnamefont {T.}~\bibnamefont
  {Egami}}, \bibinfo {author} {\bibfnamefont {S.}~\bibnamefont {Ishihara}}, \
  and\ \bibinfo {author} {\bibfnamefont {M.}~\bibnamefont {Tachiki}},\ }\href
  {\doibase 10.1126/science.261.5126.1307} {\bibfield  {journal} {\bibinfo
  {journal} {Science}\ }\textbf {\bibinfo {volume} {261}},\ \bibinfo {pages}
    {1307} (\bibinfo {year} {1993})}
  \BibitemShut
  {NoStop}%
\bibitem [{\citenamefont {Fabrizio}\ \emph {et~al.}(1999)\citenamefont
  {Fabrizio}, \citenamefont {Gogolin},\ and\ \citenamefont
  {Nersesyan}}]{FabrizioNersesyan1999}%
  \BibitemOpen
  \bibfield  {author} {\bibinfo {author} {\bibfnamefont {M.}~\bibnamefont
  {Fabrizio}}, \bibinfo {author} {\bibfnamefont {A.~O.}\ \bibnamefont
  {Gogolin}}, \ and\ \bibinfo {author} {\bibfnamefont {A.~A.}\ \bibnamefont
  {Nersesyan}},\ }\href {\doibase 10.1103/PhysRevLett.83.2014} {\bibfield
  {journal} {\bibinfo  {journal} {Phys. Rev. Lett.}\ }\textbf {\bibinfo
  {volume} {83}},\ \bibinfo {pages} {2014} (\bibinfo {year}
  {1999})}\BibitemShut {NoStop}%
\bibitem [{\citenamefont {Fabrizio}\ \emph {et~al.}(2000)\citenamefont
  {Fabrizio}, \citenamefont {Gogolin},\ and\ \citenamefont
  {Nersesyan}}]{FabrizioNersesyan2000}%
  \BibitemOpen
  \bibfield  {author} {\bibinfo {author} {\bibfnamefont {M.}~\bibnamefont
  {Fabrizio}}, \bibinfo {author} {\bibfnamefont {A.}~\bibnamefont {Gogolin}}, \
  and\ \bibinfo {author} {\bibfnamefont {A.}~\bibnamefont {Nersesyan}},\ }\href
  {\doibase http://dx.doi.org/10.1016/S0550-3213(00)00247-9} {\bibfield
  {journal} {\bibinfo  {journal} {Nuclear Physics B}\ }\textbf {\bibinfo
  {volume} {580}},\ \bibinfo {pages} {647 } (\bibinfo {year}
  {2000})}\BibitemShut {NoStop}%
\bibitem [{\citenamefont {{Gidopoulos, N.}}\ \emph {et~al.}(2000)\citenamefont
  {{Gidopoulos, N.}}, \citenamefont {{Sorella, S.}},\ and\ \citenamefont
  {{Tosatti, E.}}}]{GidopoulosTosatti2000}%
  \BibitemOpen
  \bibfield  {author} {\bibinfo {author} {\bibnamefont {{N. Gidopoulos}}},
  \bibinfo {author} {\bibnamefont {{S. Sorella}}}, \ and\ \bibinfo {author}
  {\bibnamefont {{E. Tosatti}}},\ }\href {\doibase 10.1007/s100510050123}
  {\bibfield  {journal} {\bibinfo  {journal} {Eur. Phys. J. B}\ }\textbf
  {\bibinfo {volume} {14}},\ \bibinfo {pages} {217} (\bibinfo {year}
  {2000})}\BibitemShut {NoStop}%
\bibitem [{\citenamefont {Wilkens}\ and\ \citenamefont
  {Martin}(2001)}]{WilkensMartin2001}%
  \BibitemOpen
  \bibfield  {author} {\bibinfo {author} {\bibfnamefont {T.}~\bibnamefont
  {Wilkens}}\ and\ \bibinfo {author} {\bibfnamefont {R.~M.}\ \bibnamefont
  {Martin}},\ }\href {\doibase 10.1103/PhysRevB.63.235108} {\bibfield
  {journal} {\bibinfo  {journal} {Phys. Rev. B}\ }\textbf {\bibinfo {volume}
  {63}},\ \bibinfo {pages} {235108} (\bibinfo {year} {2001})}\BibitemShut
  {NoStop}%
\bibitem [{\citenamefont {Torio}\ \emph {et~al.}(2001)\citenamefont {Torio},
  \citenamefont {Aligia},\ and\ \citenamefont {Ceccatto}}]{TorioCeccatto2001}%
  \BibitemOpen
  \bibfield  {author} {\bibinfo {author} {\bibfnamefont {M.~E.}\ \bibnamefont
  {Torio}}, \bibinfo {author} {\bibfnamefont {A.~A.}\ \bibnamefont {Aligia}}, \
  and\ \bibinfo {author} {\bibfnamefont {H.~A.}\ \bibnamefont {Ceccatto}},\
  }\href {\doibase 10.1103/PhysRevB.64.121105} {\bibfield  {journal} {\bibinfo
  {journal} {Phys. Rev. B}\ }\textbf {\bibinfo {volume} {64}},\ \bibinfo
  {pages} {121105} (\bibinfo {year} {2001})}\BibitemShut {NoStop}%
\bibitem [{\citenamefont {Kampf}\ \emph {et~al.}(2003)\citenamefont {Kampf},
  \citenamefont {Sekania}, \citenamefont {Japaridze},\ and\ \citenamefont
  {Brune}}]{KampfBrune2003}%
  \BibitemOpen
  \bibfield  {author} {\bibinfo {author} {\bibfnamefont {A.~P.}\ \bibnamefont
  {Kampf}}, \bibinfo {author} {\bibfnamefont {M.}~\bibnamefont {Sekania}},
  \bibinfo {author} {\bibfnamefont {G.~I.}\ \bibnamefont {Japaridze}}, \ and\
  \bibinfo {author} {\bibfnamefont {P.}~\bibnamefont {Brune}},\ }\href
  {http://stacks.iop.org/0953-8984/15/i=34/a=319} {\bibfield  {journal}
  {\bibinfo  {journal} {Journal of Physics: Condensed Matter}\ }\textbf
  {\bibinfo {volume} {15}},\ \bibinfo {pages} {5895} (\bibinfo {year}
  {2003})}\BibitemShut {NoStop}%
\bibitem [{\citenamefont {Zhang}\ \emph {et~al.}(2003)\citenamefont {Zhang},
  \citenamefont {Wu},\ and\ \citenamefont {Lin}}]{ZhangLin2003}%
  \BibitemOpen
  \bibfield  {author} {\bibinfo {author} {\bibfnamefont {Y.~Z.}\ \bibnamefont
  {Zhang}}, \bibinfo {author} {\bibfnamefont {C.~Q.}\ \bibnamefont {Wu}}, \
  and\ \bibinfo {author} {\bibfnamefont {H.~Q.}\ \bibnamefont {Lin}},\ }\href
  {\doibase 10.1103/PhysRevB.67.205109} {\bibfield  {journal} {\bibinfo
  {journal} {Phys. Rev. B}\ }\textbf {\bibinfo {volume} {67}},\ \bibinfo
  {pages} {205109} (\bibinfo {year} {2003})}\BibitemShut {NoStop}%
\bibitem [{\citenamefont {Batista}\ and\ \citenamefont
  {Aligia}(2004)}]{BatistaAligia2004}%
  \BibitemOpen
  \bibfield  {author} {\bibinfo {author} {\bibfnamefont {C.~D.}\ \bibnamefont
  {Batista}}\ and\ \bibinfo {author} {\bibfnamefont {A.~A.}\ \bibnamefont
  {Aligia}},\ }\href {\doibase 10.1103/PhysRevLett.92.246405} {\bibfield
  {journal} {\bibinfo  {journal} {Phys. Rev. Lett.}\ }\textbf {\bibinfo
  {volume} {92}},\ \bibinfo {pages} {246405} (\bibinfo {year}
  {2004})}\BibitemShut {NoStop}%
\bibitem [{\citenamefont {Manmana}\ \emph {et~al.}(2004)\citenamefont
  {Manmana}, \citenamefont {Meden}, \citenamefont {Noack},\ and\ \citenamefont
  {Sch\"onhammer}}]{ManmanaSchoenhammer2004}%
  \BibitemOpen
  \bibfield  {author} {\bibinfo {author} {\bibfnamefont {S.~R.}\ \bibnamefont
  {Manmana}}, \bibinfo {author} {\bibfnamefont {V.}~\bibnamefont {Meden}},
  \bibinfo {author} {\bibfnamefont {R.~M.}\ \bibnamefont {Noack}}, \ and\
  \bibinfo {author} {\bibfnamefont {K.}~\bibnamefont {Sch\"onhammer}},\ }\href
  {\doibase 10.1103/PhysRevB.70.155115} {\bibfield  {journal} {\bibinfo
  {journal} {Phys. Rev. B}\ }\textbf {\bibinfo {volume} {70}},\ \bibinfo
  {pages} {155115} (\bibinfo {year} {2004})}\BibitemShut {NoStop}%
\bibitem [{\citenamefont {Otsuka}\ and\ \citenamefont
  {Nakamura}(2005)}]{OtsukaNakamura2005}%
  \BibitemOpen
  \bibfield  {author} {\bibinfo {author} {\bibfnamefont {H.}~\bibnamefont
  {Otsuka}}\ and\ \bibinfo {author} {\bibfnamefont {M.}~\bibnamefont
  {Nakamura}},\ }\href {\doibase 10.1103/PhysRevB.71.155105} {\bibfield
  {journal} {\bibinfo  {journal} {Phys. Rev. B}\ }\textbf {\bibinfo {volume}
  {71}},\ \bibinfo {pages} {155105} (\bibinfo {year} {2005})}\BibitemShut
  {NoStop}%
\bibitem [{\citenamefont {Legeza}\ \emph {et~al.}(2006)\citenamefont {Legeza},
  \citenamefont {Buchta},\ and\ \citenamefont {S\'olyom}}]{LegezaSolyom2006}%
  \BibitemOpen
  \bibfield  {author} {\bibinfo {author} {\bibfnamefont {O.}~\bibnamefont
  {Legeza}}, \bibinfo {author} {\bibfnamefont {K.}~\bibnamefont {Buchta}}, \
  and\ \bibinfo {author} {\bibfnamefont {J.}~\bibnamefont {S\'olyom}},\ }\href
  {\doibase 10.1103/PhysRevB.73.165124} {\bibfield  {journal} {\bibinfo
  {journal} {Phys. Rev. B}\ }\textbf {\bibinfo {volume} {73}},\ \bibinfo
  {pages} {165124} (\bibinfo {year} {2006})}\BibitemShut {NoStop}%
\bibitem [{\citenamefont {Tincani}\ \emph {et~al.}(2009)\citenamefont
  {Tincani}, \citenamefont {Noack},\ and\ \citenamefont
  {Baeriswyl}}]{TincaniBaeriswyl2009}%
  \BibitemOpen
  \bibfield  {author} {\bibinfo {author} {\bibfnamefont {L.}~\bibnamefont
  {Tincani}}, \bibinfo {author} {\bibfnamefont {R.~M.}\ \bibnamefont {Noack}},
  \ and\ \bibinfo {author} {\bibfnamefont {D.}~\bibnamefont {Baeriswyl}},\
  }\href {\doibase 10.1103/PhysRevB.79.165109} {\bibfield  {journal} {\bibinfo
  {journal} {Phys. Rev. B}\ }\textbf {\bibinfo {volume} {79}},\ \bibinfo
  {pages} {165109} (\bibinfo {year} {2009})}\BibitemShut {NoStop}%
\bibitem [{\citenamefont {Aligia}\ and\ \citenamefont
  {Batista}(2005)}]{AligiaBatista2005}%
  \BibitemOpen
  \bibfield  {author} {\bibinfo {author} {\bibfnamefont {A.~A.}\ \bibnamefont
  {Aligia}}\ and\ \bibinfo {author} {\bibfnamefont {C.~D.}\ \bibnamefont
  {Batista}},\ }\href {\doibase 10.1103/PhysRevB.71.125110} {\bibfield
  {journal} {\bibinfo  {journal} {Phys. Rev. B}\ }\textbf {\bibinfo {volume}
  {71}},\ \bibinfo {pages} {125110} (\bibinfo {year} {2005})}\BibitemShut
  {NoStop}%
\bibitem [{\citenamefont {Hafez}\ and\ \citenamefont
  {Jafari}(2010)}]{HafezJafari2010}%
  \BibitemOpen
  \bibfield  {author} {\bibinfo {author} {\bibfnamefont {M.}~\bibnamefont
  {Hafez}}\ and\ \bibinfo {author} {\bibfnamefont {S.~A.}\ \bibnamefont
  {Jafari}},\ }\href@noop {} {\bibfield  {journal} {\bibinfo  {journal}
  {arXiv:1004.4265}\ } (\bibinfo {year} {2010})}\BibitemShut {NoStop}%
\bibitem [{\citenamefont {Go}\ and\ \citenamefont {Jeon}(2011)}]{GoJeon2011}%
  \BibitemOpen
  \bibfield  {author} {\bibinfo {author} {\bibfnamefont {A.}~\bibnamefont
  {Go}}\ and\ \bibinfo {author} {\bibfnamefont {G.~S.}\ \bibnamefont {Jeon}},\
  }\href {\doibase 10.1103/PhysRevB.84.195102} {\bibfield  {journal} {\bibinfo
  {journal} {Phys. Rev. B}\ }\textbf {\bibinfo {volume} {84}},\ \bibinfo
  {pages} {195102} (\bibinfo {year} {2011})}\BibitemShut {NoStop}%
\bibitem [{\citenamefont {Hafez~Torbati}\ \emph {et~al.}(2014)\citenamefont
  {Hafez~Torbati}, \citenamefont {Drescher},\ and\ \citenamefont
  {Uhrig}}]{HafezTorbatiUhrig2014}%
  \BibitemOpen
  \bibfield  {author} {\bibinfo {author} {\bibfnamefont {M.}~\bibnamefont
  {Hafez-Torbati}}, \bibinfo {author} {\bibfnamefont {N.~A.}\ \bibnamefont
  {Drescher}}, \ and\ \bibinfo {author} {\bibfnamefont {G.~S.}\ \bibnamefont
  {Uhrig}},\ }\href {\doibase 10.1103/PhysRevB.89.245126} {\bibfield  {journal}
  {\bibinfo  {journal} {Phys. Rev. B}\ }\textbf {\bibinfo {volume} {89}},\
  \bibinfo {pages} {245126} (\bibinfo {year} {2014})}\BibitemShut {NoStop}%
\bibitem [{\citenamefont {{Hafez-Torbati, Mohsen}}\ \emph
  {et~al.}(2015)\citenamefont {{Hafez-Torbati, Mohsen}}, \citenamefont
  {{Drescher, Nils A.}},\ and\ \citenamefont {{Uhrig, Götz
  S.}}}]{HafezTorbatiUhrig2015}%
  \BibitemOpen
  \bibfield  {author} {\bibinfo {author} {\bibnamefont {{M. Hafez-Torbati}}},
    \bibinfo {author} {\bibnamefont {{N.~A. Drescher}}}, \ and\
  \bibinfo {author} {\bibnamefont {{G.~S. Uhrig}}},\ }\href {\doibase
  10.1140/epjb/e2014-50551-0} {\bibfield  {journal} {\bibinfo  {journal} {Eur.
  Phys. J. B}\ }\textbf {\bibinfo {volume} {88}},\ \bibinfo {pages} {36}
  (\bibinfo {year} {2015})}\BibitemShut {NoStop}%
  %
  %
  %
\bibitem{footnote1} It should be emphasized that the amplitude of the modulation $A$, here $A=0.005J$,
  needs to be chosen carefully. For small frequencies the period $T = 2\pi/\omega$ of the modulation
  becomes large such that one can only observe a few periods within a realistic time interval.
  For amplitudes smaller than $0.005J$, the absorption onset shifts to later times: measuring the
  absorption rate then becomes even more difficult.
  %
  %
  %
\bibitem [{\citenamefont {Tarruell}\ \emph {et~al.}(2012)\citenamefont
  {Tarruell}, \citenamefont {Greif}, \citenamefont {Uehlinger}, \citenamefont
  {Jotzu},\ and\ \citenamefont {Esslinger}}]{TarruellEsslinger2012}%
  \BibitemOpen
  \bibfield  {author} {\bibinfo {author} {\bibfnamefont {L.}~\bibnamefont
  {Tarruell}}, \bibinfo {author} {\bibfnamefont {D.}~\bibnamefont {Greif}},
  \bibinfo {author} {\bibfnamefont {T.}~\bibnamefont {Uehlinger}}, \bibinfo
  {author} {\bibfnamefont {G.}~\bibnamefont {Jotzu}}, \ and\ \bibinfo {author}
  {\bibfnamefont {T.}~\bibnamefont {Esslinger}},\ }\href {\doibase
  10.1038/nature10871} {\bibfield  {journal} {\bibinfo  {journal} {Nature}\
  }\textbf {\bibinfo {volume} {483}},\ \bibinfo {pages} {302} (\bibinfo {year}
  {2012})}\BibitemShut {NoStop}%
\bibitem [{\citenamefont {Bloch}\ \emph {et~al.}(2008)\citenamefont {Bloch},
  \citenamefont {Dalibard},\ and\ \citenamefont {Zwerger}}]{BlochZwerger2008}%
  \BibitemOpen
  \bibfield  {author} {\bibinfo {author} {\bibfnamefont {I.}~\bibnamefont
  {Bloch}}, \bibinfo {author} {\bibfnamefont {J.}~\bibnamefont {Dalibard}}, \
  and\ \bibinfo {author} {\bibfnamefont {W.}~\bibnamefont {Zwerger}},\ }\href
  {\doibase 10.1103/RevModPhys.80.885} {\bibfield  {journal} {\bibinfo
  {journal} {Rev. Mod. Phys.}\ }\textbf {\bibinfo {volume} {80}},\ \bibinfo
  {pages} {885} (\bibinfo {year} {2008})}\BibitemShut {NoStop}%
\bibitem{ToermaeTarruell2015}%
  \BibitemOpen
  \bibfield  {author} {\bibinfo {author} {\bibfnamefont {P. T\"orm\"a and L. Tarruel}}}
  in\ \href@noop {} {\emph {\bibinfo {booktitle} {Quantum Gas
  Experiments}}},\ \bibinfo {editor} {edited by\ \bibinfo {editor}
  {\bibfnamefont {P.}~\bibnamefont {T\"orm\"a}}\ and\ \bibinfo {editor}
  {\bibfnamefont {K.}~\bibnamefont {Sengstock}}}\ (\bibinfo  {publisher}
  {Imperial College Press},\ \bibinfo {address} {London},\ \bibinfo {year}
  {2015})\ Chap.~\bibinfo {chapter} {10-11}, pp.\ \bibinfo {pages}
  {199--266}\BibitemShut {NoStop}%
\bibitem [{\citenamefont {Kollath}\ \emph {et~al.}(2006)\citenamefont
  {Kollath}, \citenamefont {Iucci}, \citenamefont {McCulloch},\ and\
  \citenamefont {Giamarchi}}]{KollathGiamarchi2006}%
  \BibitemOpen
  \bibfield  {author} {\bibinfo {author} {\bibfnamefont {C.}~\bibnamefont
  {Kollath}}, \bibinfo {author} {\bibfnamefont {A.}~\bibnamefont {Iucci}},
  \bibinfo {author} {\bibfnamefont {I.~P.}\ \bibnamefont {McCulloch}}, \ and\
  \bibinfo {author} {\bibfnamefont {T.}~\bibnamefont {Giamarchi}},\ }\href@noop
  {} {\bibfield  {journal} {\bibinfo  {journal} {Phys. Rev. A}\ }\textbf
  {\bibinfo {volume} {74}},\ \bibinfo {pages} {041604} (\bibinfo {year}
  {2006})}\BibitemShut {NoStop}%
\bibitem [{\citenamefont {Loida}\ \emph {et~al.}(2015)\citenamefont {Loida},
  \citenamefont {Sheikhan},\ and\ \citenamefont {Kollath}}]{LoidaKollath2015}%
  \BibitemOpen
  \bibfield  {author} {\bibinfo {author} {\bibfnamefont {K.}~\bibnamefont
  {Loida}}, \bibinfo {author} {\bibfnamefont {A.}~\bibnamefont {Sheikhan}}, \
  and\ \bibinfo {author} {\bibfnamefont {C.}~\bibnamefont {Kollath}},\ }\href
  {\doibase 10.1103/PhysRevA.92.043624} {\bibfield  {journal} {\bibinfo
  {journal} {Phys. Rev. A}\ }\textbf {\bibinfo {volume} {92}},\ \bibinfo
  {pages} {043624} (\bibinfo {year} {2015})}\BibitemShut {NoStop}%
\bibitem [{\citenamefont {Huber}\ and\ \citenamefont
  {R\"uegg}(2009)}]{HuberRuegg2009}%
  \BibitemOpen
  \bibfield  {author} {\bibinfo {author} {\bibfnamefont {S.~D.}\ \bibnamefont
  {Huber}}\ and\ \bibinfo {author} {\bibfnamefont {A.}~\bibnamefont
  {R\"uegg}},\ }\href {\doibase 10.1103/PhysRevLett.102.065301} {\bibfield
  {journal} {\bibinfo  {journal} {Phys. Rev. Lett.}\ }\textbf {\bibinfo
  {volume} {102}},\ \bibinfo {pages} {065301} (\bibinfo {year}
  {2009})}\BibitemShut {NoStop}%
\bibitem [{\citenamefont {Sensarma}\ \emph {et~al.}(2009)\citenamefont
  {Sensarma}, \citenamefont {Pekker}, \citenamefont {Lukin},\ and\
  \citenamefont {Demler}}]{SensarmaDemler2009}%
  \BibitemOpen
  \bibfield  {author} {\bibinfo {author} {\bibfnamefont {R.}~\bibnamefont
  {Sensarma}}, \bibinfo {author} {\bibfnamefont {D.}~\bibnamefont {Pekker}},
  \bibinfo {author} {\bibfnamefont {M.~D.}\ \bibnamefont {Lukin}}, \ and\
  \bibinfo {author} {\bibfnamefont {E.}~\bibnamefont {Demler}},\ }\href
  {\doibase 10.1103/PhysRevLett.103.035303} {\bibfield  {journal} {\bibinfo
  {journal} {Phys. Rev. Lett.}\ }\textbf {\bibinfo {volume} {103}},\ \bibinfo
  {pages} {035303} (\bibinfo {year} {2009})}\BibitemShut {NoStop}%
\bibitem [{\citenamefont {Massel}\ \emph {et~al.}(2009)\citenamefont {Massel},
  \citenamefont {Leskinen},\ and\ \citenamefont
  {T\"orm\"a}}]{MasselToermae2009}%
  \BibitemOpen
  \bibfield  {author} {\bibinfo {author} {\bibfnamefont {F.}~\bibnamefont
  {Massel}}, \bibinfo {author} {\bibfnamefont {M.~J.}\ \bibnamefont
  {Leskinen}}, \ and\ \bibinfo {author} {\bibfnamefont {P.}~\bibnamefont
  {T\"orm\"a}},\ }\href {\doibase 10.1103/PhysRevLett.103.066404} {\bibfield
  {journal} {\bibinfo  {journal} {Phys. Rev. Lett.}\ }\textbf {\bibinfo
  {volume} {103}},\ \bibinfo {pages} {066404} (\bibinfo {year}
  {2009})}\BibitemShut {NoStop}%
\bibitem [{\citenamefont {Korolyuk}\ \emph {et~al.}(2010)\citenamefont
  {Korolyuk}, \citenamefont {Massel},\ and\ \citenamefont
  {T\"orm\"a}}]{KorolyukToermae2010}%
  \BibitemOpen
  \bibfield  {author} {\bibinfo {author} {\bibfnamefont {A.}~\bibnamefont
  {Korolyuk}}, \bibinfo {author} {\bibfnamefont {F.}~\bibnamefont {Massel}}, \
  and\ \bibinfo {author} {\bibfnamefont {P.}~\bibnamefont {T\"orm\"a}},\ }\href
  {\doibase 10.1103/PhysRevLett.104.236402} {\bibfield  {journal} {\bibinfo
  {journal} {Phys. Rev. Lett.}\ }\textbf {\bibinfo {volume} {104}},\ \bibinfo
  {pages} {236402} (\bibinfo {year} {2010})}\BibitemShut {NoStop}%
\bibitem [{\citenamefont {Xu}\ \emph {et~al.}(2011)\citenamefont {Xu},
  \citenamefont {Chiesa}, \citenamefont {Yang}, \citenamefont {Su},
  \citenamefont {Sheehy}, \citenamefont {Moreno}, \citenamefont {Scalettar},\
  and\ \citenamefont {Jarrell}}]{XuJarrell2011}%
  \BibitemOpen
  \bibfield  {author} {\bibinfo {author} {\bibfnamefont {Z.}~\bibnamefont
  {Xu}}, \bibinfo {author} {\bibfnamefont {S.}~\bibnamefont {Chiesa}}, \bibinfo
  {author} {\bibfnamefont {S.}~\bibnamefont {Yang}}, \bibinfo {author}
  {\bibfnamefont {S.-Q.}\ \bibnamefont {Su}}, \bibinfo {author} {\bibfnamefont
  {D.~E.}\ \bibnamefont {Sheehy}}, \bibinfo {author} {\bibfnamefont
  {J.}~\bibnamefont {Moreno}}, \bibinfo {author} {\bibfnamefont {R.~T.}\
  \bibnamefont {Scalettar}}, \ and\ \bibinfo {author} {\bibfnamefont
  {M.}~\bibnamefont {Jarrell}},\ }\href {\doibase 10.1103/PhysRevA.84.021607}
  {\bibfield  {journal} {\bibinfo  {journal} {Phys. Rev. A}\ }\textbf {\bibinfo
  {volume} {84}},\ \bibinfo {pages} {021607} (\bibinfo {year}
  {2011})}\BibitemShut {NoStop}%
\bibitem [{\citenamefont {Tokuno}\ \emph {et~al.}(2012)\citenamefont {Tokuno},
  \citenamefont {Demler},\ and\ \citenamefont
  {Giamarchi}}]{TokunoGiamarchi2012_0}%
  \BibitemOpen
  \bibfield  {author} {\bibinfo {author} {\bibfnamefont {A.}~\bibnamefont
  {Tokuno}}, \bibinfo {author} {\bibfnamefont {E.}~\bibnamefont {Demler}}, \
  and\ \bibinfo {author} {\bibfnamefont {T.}~\bibnamefont {Giamarchi}},\ }\href
  {\doibase 10.1103/PhysRevA.85.053601} {\bibfield  {journal} {\bibinfo
  {journal} {Phys. Rev. A}\ }\textbf {\bibinfo {volume} {85}},\ \bibinfo
  {pages} {053601} (\bibinfo {year} {2012})}\BibitemShut {NoStop}%
\bibitem [{\citenamefont {Tokuno}\ and\ \citenamefont
  {Giamarchi}(2012)}]{TokunoGiamarchi2012}%
  \BibitemOpen
  \bibfield  {author} {\bibinfo {author} {\bibfnamefont {A.}~\bibnamefont
  {Tokuno}}\ and\ \bibinfo {author} {\bibfnamefont {T.}~\bibnamefont
  {Giamarchi}},\ }\href {\doibase 10.1103/PhysRevA.85.061603} {\bibfield
  {journal} {\bibinfo  {journal} {Phys. Rev. A}\ }\textbf {\bibinfo {volume}
  {85}},\ \bibinfo {pages} {061603} (\bibinfo {year} {2012})}\BibitemShut
  {NoStop}%
\bibitem [{\citenamefont {Dirks}\ \emph {et~al.}(2014)\citenamefont {Dirks},
  \citenamefont {Mikelsons}, \citenamefont {Krishnamurthy},\ and\ \citenamefont
  {Freericks}}]{DirksFreericks2014}%
  \BibitemOpen
  \bibfield  {author} {\bibinfo {author} {\bibfnamefont {A.}~\bibnamefont
  {Dirks}}, \bibinfo {author} {\bibfnamefont {K.}~\bibnamefont {Mikelsons}},
  \bibinfo {author} {\bibfnamefont {H.~R.}\ \bibnamefont {Krishnamurthy}}, \
  and\ \bibinfo {author} {\bibfnamefont {J.~K.}\ \bibnamefont {Freericks}},\
  }\href {\doibase 10.1103/PhysRevA.89.021602} {\bibfield  {journal} {\bibinfo
  {journal} {Phys. Rev. A}\ }\textbf {\bibinfo {volume} {89}},\ \bibinfo
  {pages} {021602} (\bibinfo {year} {2014})}\BibitemShut {NoStop}%
\bibitem [{\citenamefont {J\"ordens}\ \emph {et~al.}(2008)\citenamefont
  {J\"ordens}, \citenamefont {Strohmaier}, \citenamefont {G\"unter},
  \citenamefont {Moritz},\ and\ \citenamefont
  {Esslinger}}]{JoerdensEsslinger2008}%
  \BibitemOpen
  \bibfield  {author} {\bibinfo {author} {\bibfnamefont {R.}~\bibnamefont
  {J\"ordens}}, \bibinfo {author} {\bibfnamefont {N.}~\bibnamefont
  {Strohmaier}}, \bibinfo {author} {\bibfnamefont {K.}~\bibnamefont
  {G\"unter}}, \bibinfo {author} {\bibfnamefont {H.}~\bibnamefont {Moritz}}, \
  and\ \bibinfo {author} {\bibfnamefont {T.}~\bibnamefont {Esslinger}},\
  }\href@noop {} {\bibfield  {journal} {\bibinfo  {journal} {Nature}\ }\textbf
  {\bibinfo {volume} {455}},\ \bibinfo {pages} {204} (\bibinfo {year}
  {2008})}\BibitemShut {NoStop}%
\bibitem [{\citenamefont {Greif}\ \emph {et~al.}(2011)\citenamefont {Greif},
  \citenamefont {Tarruell}, \citenamefont {Uehlinger}, \citenamefont
  {J\"ordens},\ and\ \citenamefont {Esslinger}}]{GreifEsslinger2011}%
  \BibitemOpen
  \bibfield  {author} {\bibinfo {author} {\bibfnamefont {D.}~\bibnamefont
  {Greif}}, \bibinfo {author} {\bibfnamefont {L.}~\bibnamefont {Tarruell}},
  \bibinfo {author} {\bibfnamefont {T.}~\bibnamefont {Uehlinger}}, \bibinfo
  {author} {\bibfnamefont {R.}~\bibnamefont {J\"ordens}}, \ and\ \bibinfo
  {author} {\bibfnamefont {T.}~\bibnamefont {Esslinger}},\ }\href {\doibase
  10.1103/PhysRevLett.106.145302} {\bibfield  {journal} {\bibinfo  {journal}
  {Phys. Rev. Lett.}\ }\textbf {\bibinfo {volume} {106}},\ \bibinfo {pages}
  {145302} (\bibinfo {year} {2011})}\BibitemShut {NoStop}%
\bibitem [{\citenamefont {Strohmaier}\ \emph {et~al.}(2010)\citenamefont
  {Strohmaier}, \citenamefont {Greif}, \citenamefont {J\"ordens}, \citenamefont
  {Tarruell}, \citenamefont {Moritz}, \citenamefont {Esslinger}, \citenamefont
  {Sensarma}, \citenamefont {Pekker}, \citenamefont {Altman},\ and\
  \citenamefont {Demler}}]{StrohmaierDemler2010}%
  \BibitemOpen
  \bibfield  {author} {\bibinfo {author} {\bibfnamefont {N.}~\bibnamefont
  {Strohmaier}}, \bibinfo {author} {\bibfnamefont {D.}~\bibnamefont {Greif}},
  \bibinfo {author} {\bibfnamefont {R.}~\bibnamefont {J\"ordens}}, \bibinfo
  {author} {\bibfnamefont {L.}~\bibnamefont {Tarruell}}, \bibinfo {author}
  {\bibfnamefont {H.}~\bibnamefont {Moritz}}, \bibinfo {author} {\bibfnamefont
  {T.}~\bibnamefont {Esslinger}}, \bibinfo {author} {\bibfnamefont
  {R.}~\bibnamefont {Sensarma}}, \bibinfo {author} {\bibfnamefont
  {D.}~\bibnamefont {Pekker}}, \bibinfo {author} {\bibfnamefont
  {E.}~\bibnamefont {Altman}}, \ and\ \bibinfo {author} {\bibfnamefont
  {E.}~\bibnamefont {Demler}},\ }\href@noop {} {\bibfield  {journal} {\bibinfo
  {journal} {Phys.~ Rev.~ Lett.}\ }\textbf {\bibinfo {volume} {104}},\ \bibinfo
  {pages} {080401} (\bibinfo {year} {2010})}\BibitemShut {NoStop}%
\bibitem [{\citenamefont {Heinze}\ \emph {et~al.}(2011)\citenamefont {Heinze},
  \citenamefont {G\"otze}, \citenamefont {Krauser}, \citenamefont {Hundt},
  \citenamefont {Fl\"aschner}, \citenamefont {L\"uhmann}, \citenamefont
  {Becker},\ and\ \citenamefont {Sengstock}}]{HeinzeSengstock2011}%
  \BibitemOpen
  \bibfield  {author} {\bibinfo {author} {\bibfnamefont {J.}~\bibnamefont
  {Heinze}}, \bibinfo {author} {\bibfnamefont {S.}~\bibnamefont {G\"otze}},
  \bibinfo {author} {\bibfnamefont {J.~S.}\ \bibnamefont {Krauser}}, \bibinfo
  {author} {\bibfnamefont {B.}~\bibnamefont {Hundt}}, \bibinfo {author}
  {\bibfnamefont {N.}~\bibnamefont {Fl\"aschner}}, \bibinfo {author}
  {\bibfnamefont {D.-S.}\ \bibnamefont {L\"uhmann}}, \bibinfo {author}
  {\bibfnamefont {C.}~\bibnamefont {Becker}}, \ and\ \bibinfo {author}
  {\bibfnamefont {K.}~\bibnamefont {Sengstock}},\ }\href {\doibase
  10.1103/PhysRevLett.107.135303} {\bibfield  {journal} {\bibinfo  {journal}
  {Phys. Rev. Lett.}\ }\textbf {\bibinfo {volume} {107}},\ \bibinfo {pages}
  {135303} (\bibinfo {year} {2011})}\BibitemShut {NoStop}%
\bibitem [{\citenamefont {Heinze}\ \emph {et~al.}(2013)\citenamefont {Heinze},
  \citenamefont {Krauser}, \citenamefont {Fl\"aschner}, \citenamefont {Hundt},
  \citenamefont {G\"otze}, \citenamefont {Itin}, \citenamefont {Mathey},
  \citenamefont {Sengstock},\ and\ \citenamefont {Becker}}]{HeinzeBecker2013}%
  \BibitemOpen
  \bibfield  {author} {\bibinfo {author} {\bibfnamefont {J.}~\bibnamefont
  {Heinze}}, \bibinfo {author} {\bibfnamefont {J.~S.}\ \bibnamefont {Krauser}},
  \bibinfo {author} {\bibfnamefont {N.}~\bibnamefont {Fl\"aschner}}, \bibinfo
  {author} {\bibfnamefont {B.}~\bibnamefont {Hundt}}, \bibinfo {author}
  {\bibfnamefont {S.}~\bibnamefont {G\"otze}}, \bibinfo {author} {\bibfnamefont
  {A.~P.}\ \bibnamefont {Itin}}, \bibinfo {author} {\bibfnamefont
  {L.}~\bibnamefont {Mathey}}, \bibinfo {author} {\bibfnamefont
  {K.}~\bibnamefont {Sengstock}}, \ and\ \bibinfo {author} {\bibfnamefont
  {C.}~\bibnamefont {Becker}},\ }\href {\doibase
  10.1103/PhysRevLett.110.085302} {\bibfield  {journal} {\bibinfo  {journal}
  {Phys. Rev. Lett.}\ }\textbf {\bibinfo {volume} {110}},\ \bibinfo {pages}
  {085302} (\bibinfo {year} {2013})}\BibitemShut {NoStop}%
\bibitem [{\citenamefont {Messer}\ \emph {et~al.}(2015)\citenamefont {Messer},
  \citenamefont {Desbuquois}, \citenamefont {Uehlinger}, \citenamefont {Jotzu},
  \citenamefont {Huber}, \citenamefont {Greif},\ and\ \citenamefont
  {Esslinger}}]{MesserEsslinger2015}%
  \BibitemOpen
  \bibfield  {author} {\bibinfo {author} {\bibfnamefont {M.}~\bibnamefont
  {Messer}}, \bibinfo {author} {\bibfnamefont {R.}~\bibnamefont {Desbuquois}},
  \bibinfo {author} {\bibfnamefont {T.}~\bibnamefont {Uehlinger}}, \bibinfo
  {author} {\bibfnamefont {G.}~\bibnamefont {Jotzu}}, \bibinfo {author}
  {\bibfnamefont {S.}~\bibnamefont {Huber}}, \bibinfo {author} {\bibfnamefont
  {D.}~\bibnamefont {Greif}}, \ and\ \bibinfo {author} {\bibfnamefont
  {T.}~\bibnamefont {Esslinger}},\ }\href {\doibase
  10.1103/PhysRevLett.115.115303} {\bibfield  {journal} {\bibinfo  {journal}
  {Phys. Rev. Lett.}\ }\textbf {\bibinfo {volume} {115}},\ \bibinfo {pages}
    {115303} (\bibinfo {year} {2015})}\BibitemShut {NoStop}%
%
\bibitem{DaleyVidal2004} A.~J. Daley, C.~Kollath, U.~Schollw\"ock, G.~Vidal, J. Stat. Mech.: Theor. Exp. (2004) P04005.
%
\bibitem{WhiteFeiguin2004} S.~R. White and A.~E. Feiguin, Phys. Rev. Lett. {\bf 93}, 076401 (2004).
%
\bibitem [{\citenamefont {Schollw\"ock}(2011)}]{Schollwoeck2011}%
  \BibitemOpen
  \bibfield  {author} {\bibinfo {author} {\bibfnamefont {U.}~\bibnamefont
  {Schollw\"ock}},\ }\href {\doibase
  http://dx.doi.org/10.1016/j.aop.2010.09.012} {\bibfield  {journal} {\bibinfo
  {journal} {Annals of Physics}\ }\textbf {\bibinfo {volume} {326}},\ \bibinfo
  {pages} {96 } (\bibinfo {year} {2011})}\BibitemShut {NoStop}%
\bibitem [{\citenamefont {Delfino}\ and\ \citenamefont
  {Mussardo}(1998)}]{DelfinoMussardo1998}%
  \BibitemOpen
  \bibfield  {author} {\bibinfo {author} {\bibfnamefont {G.}~\bibnamefont
  {Delfino}}\ and\ \bibinfo {author} {\bibfnamefont {G.}~\bibnamefont
  {Mussardo}},\ }\href {\doibase
  http://dx.doi.org/10.1016/S0550-3213(98)00063-7} {\bibfield  {journal}
  {\bibinfo  {journal} {Nuclear Physics B}\ }\textbf {\bibinfo {volume}
  {516}},\ \bibinfo {pages} {675 } (\bibinfo {year} {1998})}\BibitemShut
  {NoStop}%
\bibitem [{\citenamefont {Bajnok}\ \emph {et~al.}(2001)\citenamefont {Bajnok},
  \citenamefont {Palla}, \citenamefont {Takács},\ and\ \citenamefont
  {Wágner}}]{BajnokWagner2001}%
  \BibitemOpen
  \bibfield  {author} {\bibinfo {author} {\bibfnamefont {Z.}~\bibnamefont
  {Bajnok}}, \bibinfo {author} {\bibfnamefont {L.}~\bibnamefont {Palla}},
  \bibinfo {author} {\bibfnamefont {G.}~\bibnamefont {Takács}}, \ and\
  \bibinfo {author} {\bibfnamefont {F.}~\bibnamefont {Wágner}},\ }\href
  {\doibase https://doi.org/10.1016/S0550-3213(01)00067-0} {\bibfield
  {journal} {\bibinfo  {journal} {Nuclear Physics B}\ }\textbf {\bibinfo
  {volume} {601}},\ \bibinfo {pages} {503 } (\bibinfo {year}
  {2001})}\BibitemShut {NoStop}%
\bibitem [{\citenamefont {Takacs}\ and\ \citenamefont
  {Wagner}(2006)}]{TakacsWagner2006}%
  \BibitemOpen
  \bibfield  {author} {\bibinfo {author} {\bibfnamefont {G.}~\bibnamefont
  {Takacs}}\ and\ \bibinfo {author} {\bibfnamefont {F.}~\bibnamefont
  {Wagner}},\ }\href {\doibase
  https://doi.org/10.1016/j.nuclphysb.2006.02.004} {\bibfield  {journal}
  {\bibinfo  {journal} {Nuclear Physics B}\ }\textbf {\bibinfo {volume}
  {741}},\ \bibinfo {pages} {353 } (\bibinfo {year} {2006})}\BibitemShut {NoStop}%
%
%
\bibitem{footnote2} In contrast, the $n=0$ peaks are expected at
  energies $\hbar \omega\sim \frac{\pi \hbar u}{4aL}$ which are too low in frequency to be resolved by our numerics.
%
%
\bibitem{inpreparation} K.~Loida, J.-S.~Bernier, R.~Citro, E.~Orignac, and C.~Kollath, in preparation. 
%
%
\bibitem [{\citenamefont {Lecheminant}(2005)}]{Lecheminant2005}%
  \BibitemOpen
  \bibfield  {author} {\bibinfo {author} {\bibfnamefont {P.}~\bibnamefont
  {Lecheminant}},\ }in\ \href@noop {} {\emph {\bibinfo {booktitle} {Frustrated
  spin systems}}},\ \bibinfo {editor} {edited by\ \bibinfo {editor}
  {\bibfnamefont {H.~T.}\ \bibnamefont {Diep}}}\ (\bibinfo  {publisher} {World
  Scientific},\ \bibinfo {address} {Singapore},\ \bibinfo {year} {2005})\
  Chap.~\bibinfo {chapter} {6}, p.\ \bibinfo {pages} {307}\BibitemShut
  {NoStop}%
\end{thebibliography}
\end{document}